# 1D van der Waals Material Tellurium: Raman Spectroscopy under Strain and Magneto-transport

Yuchen Du[1,4], Gang Qiu[1,4], Yixiu Wang[2], Mengwei Si[1,4], Xianfan Xu[3,4], Wenzhuo Wu[2,*], and Peide D. Ye[1,4,*]

[1] School of Electrical and Computer Engineering, Purdue University, West Lafayette, Indiana 47907, United States

[2] School of Industrial Engineering, Purdue University, West Lafayette, Indiana 47907, United States

[3] School of Mechanical Engineering, Purdue University, West Lafayette, Indiana 47907, United States

[4] Birck Nanotechnology Center, Purdue University, West Lafayette, Indiana 47907, United States

* Address correspondence to: yep@purdue.edu or wenzhuowu@purdue.edu

ABSTRACT

Experimental demonstrations of 1D van der Waals material tellurium have been presented by Raman spectroscopy under strain and magneto-transport. Raman spectroscopy measurements have been performed under strains along different principle axes. Pronounced strain response along *c*-axis is observed due to the strong intra-chain covalent bonds, while no strain response is obtained along *a*-axis due to the weak inter-chain van der Waals interaction. Magneto-transport results further verify its anisotropic property, resulting in dramatically distinct magneto-resistance behaviors in terms of three different magnetic field directions. Specifically, phase coherence length extracted from weak antilocalization effect, $L_\phi \sim T^{-0.5}$, claims its 2D transport characteristics when an applied magnetic field is perpendicular to the thin film. In contrast, $L_\phi \sim T^{-0.33}$ is obtained from universal conductance fluctuations once the magnetic field is along *c*-axis of Te, indicating its nature of 1D transport along the helical atomic chains. Our studies, which are obtained on high quality single crystal tellurium thin film, appear to serve as strong evidences of its 1D van der Waals structure from experimental perspectives. It is the aim of this paper to address this special concept that differs from the previous well-studied 1D nanowires or 2D van der Waals materials.

MAIN TEXT

Chalcogens, more specifically, tellurium (Te) is a *p*-type semiconductor with a bandgap of ~0.35 eV in bulk, and ~1 eV in monolayer[1-7]. The importance of introducing Te is not only based on the fact that it is an elemental metallic bonding-based piezoelectric material[8-10], but also has an asymmetric crystal structure in its radial direction. The trigonal crystalline form of Te is an array of parallel atomic chains arranged on a 2D hexagonal lattice. The packing of the chains is determined by the dashed lines in Figure 1(a), which are pure van der Waals forces in the basal plane. The solid lines designate the primary covalent bonds in the chain[11,12]. As is evident in Figure 1(a), Te has a highly anisotropic crystal structure, and it consists of helical chains along *c*-axis which are bound with other chains by van der Waals interactions. Due to its special crystal structure, the synthesis of Te is divided into two primary categories. The single 1D Te nanowire has been achieved by a variety of synthesis methods such as microwave-assisted synthesis in ionic liquid[13], vapor phase growth[14], and solution phase synthesis[15,16,17], where the growth mechanism involves the nucleation of spherical particles followed by the growth of 1D nanostructure along *c*-axis, and the diameter of the nanowires ranges from a few to hundreds of nanometers. 2D hexagonal Te nanoplates can also be realized *via* high-temperature vapor phase process[18], in which the developed Te nanoplates share the same crystal geometry as its basal unit cell, and the thickness and lateral dimension of 2D Te hexagonal nanoplate are up to several hundred nanometers. 2D Te crystals packed along *a*-axis with van der Waals forces, as shown in Figure 1(b), with Te atomic chains along primary *c*-axis, have not been realized.

Recently, we have demonstrated a substrate-free solution phase process to synthesize a new type of large-area, high-quality 2D single crystal Te films, which exhibits process-dependent thickness ranging from a few to tens of nanometers[19]. The as-fabricated Te field-effect transistors demonstrate large on/off ratio (~$10^6$), high field-effect mobilities (~700 cm$^2$/Vs), and excessive drain current of ~600 mA/mm at 100 nm channel length[19], which are comparable or superior to transistors made of other experimentally accessible 2D semiconductors. In great contrast to black phosphorus and other environmentally sensitive 2D crystals, 2D Te thin film has a pronounced air-stability. The significance of this work stands on the fact that the synthesized Te thin films have revealed mostly in trapezoid shape, where the length and width of the films can be up to ~100 µm and ~10 µm, respectively. The liquid phase synthesized Te films have a distinct crystal structure as illustrated in Figure 1(b) and 1(c) which differ from 1D Te nanowires or 2D van der Waals hexagonal Te thin film. Raman spectroscopy and transport measurements confirm that the as-synthesized Te films have all covalent 1D atomic chains along primary *c*-axis as long as 100 µm, while all helical chains are stacked together by van der Waals force along *a*-axis as wide as 10 µm.

In this letter, we have employed Raman spectroscopy to experimentally verify the 1D van der Waals structure of the Te thin films with respect to strains along different principle axes. The significant strain response along *c*-axis, together with no response along *a*-axis, has clearly demonstrated its 1D van der Waals nature and the anisotropic atomic structure. Red-shifted/blue-shifted Raman spectra of Te along *c*-axis under tensile/compressive strain can be interpreted as the result of deformation of covalent bond length in 1D helical chains, and the absence of Raman shift along strained *a*-axis reveals its weak van der Waals interaction between atomic chains. We have also carried out magneto-transport studies of Te thin films with magnetic fields along three principle axes. The measurement results further verify its anisotropic property with dramatically different characteristics. Temperature dependent phase coherence length extracted from weak antilocalization effect (WAL), $L_\phi \sim T^{-0.5}$, claims its 2D transport behavior when an applied magnetic field is perpendicular to *x-y* plane. In contrast, $L_\phi \sim T^{-0.33}$ is obtained from universal conductance fluctuations (UCF) once an applied magnetic field is along the *c*-axis of Te, indicating its nature of 1D transport mechanism. Our studies, which are obtained on high-quality single crystal Te thin films, serve as strong experimental evidence of its 1D van der Waals structure. The work presented in this letter adds a new class of 2D crystals, made of 1D van der Waals materials, to the well-studied 2D materials family.

**Crystal structure and lattice vibration of Te.** Te crystallizes in a hexagonal structure and every fourth atom is directly located above another atom in its chain, so that the projection of the chain on a plane perpendicular to the chain direction is an equilateral triangle. The hexagonal lattice is achieved by locating a chain at the center and at each of the six corners of a hexagon. Te atoms are covalently connected *via* an infinite helical turn at 120° along its longitudinal [0001] direction, which is defined as *c*-axis. All turns are bounded to hexagonal bundles in its radial direction by weak van der Waals force that has been defined as *a*-axis[11,12], whch is the in-plane direction of the 2D Te in this experiment[19]. To start our experiment, Te thin film was synthesized and transferred to a conducting Si substrate with a 300 nm $SiO_2$ capping layer[19]. A typical optical image of Te thin film is given in Figure 1(c), with a thickness of 22.1 nm determined by atomic force microscopy (AFM). Identified Te thin films have this trapezoid shape, which is easily to classify *c*-axis versus *a*-axis. Raman spectroscopy in Figure 1(d) shows Te thin film has three active Raman phonon modes. The most intense Raman peak of Te located at ~122 $cm^{-1}$ is related to the $A_1$ mode, corresponding to chain expansion mode in which each atom moves in the basal plane[12]. Meanwhile, there exist two degenerate E modes which separate into predominately bond-bending and bond-stretching types with larger admixture in Te[12]. $E^1$ mode at ~94 $cm^{-1}$ is caused by *a*-axis rotation, and $E^2$ mode at 142 cm$^{-1}$ is appointed to asymmetric stretching mainly along the *c*-axis[12].

**Significant strain-response along 1D covalent *c*-axis.** We first investigated the evolution of the Raman spectra of Te with uniaxial tensile and compressive strains, summarized in Figure 2(a). The laser polarization is aligned along the *a*-axis of Te thin film, and the strain direction is along *c*-axis based on our apparatus set-up[20]. In our work, we used Lorentzian function to fit the Raman spectra and obtained the peak frequency of each mode at different strain strengths. For unstrained Te, consistent with previous reports[3,18,21,22], we observe the $A_1$ mode at ~122 cm$^{-1}$, and modes of $E^1$ and $E^2$ at ~94 cm$^{-1}$ and ~142 cm$^{-1}$, respectively. The $A_1$ and $E^2$ modes show the same linear trend of Raman frequency shift with respect to the applied strains, while the rates of frequency shift are different in these two modes. Both $A_1$ and $E^2$ modes experience a blue-shift when Te is under compressive strains along the *c*-axis, with a slope of 1.1 cm$^{-1}$%$^{-1}$ and 0.9 cm$^{-1}$%$^{-1}$, respectively. On the other hand, $A_1$ and $E^2$ have a red-shift at a rate of 1.0 cm$^{-1}$%$^{-1}$ and 1.2 cm$^{-1}$%$^{-1}$ under uniaxial tensile strains along the *c*-axis, as shown in Figure 2(b) and (c). Note that we did not observe a measurable Raman shift in the $E^1$ mode, where the $E^1$ peak position remains the same with various *c*-axis strains and we ascribe it to the fact $E^1$ mode is responsible to *a*-axis rotation and is not sensitive to *c*-axis deformation[12,21,22]. The different responses of Raman spectra with strains can be explained by analyzing the types of vibration mode involved. Let us take the $E^2$ mode as an example in the first place, where the atomic motions of $E^2$ mode occurs along the 1D *c*-axis direction. The red-shift of the $E^2$ mode under tensile strain along the *c*-axis, with a slope of 1.2 cm$^{-1}$%$^{-1}$, can be understood based on the elongation of the Te-Te covalent bond length, which weakens the interatomic interactions and therefore reducing the vibration frequency[20,23]. Meanwhile, blue-shift with a slope of 0.9 cm$^{-1}$%$^{-1}$ under compressive strain indicates the enhancement of interatomic interaction, thus interpreted as a result of the shortened covalent Te-Te bond under compressive deformation. Furthermore, tensile strain not only enlarges the Te-Te bond length assigned to $E^2$ mode along *c*-axis, but also influences the basal plane displacement pattern which is associated with $A_1$ mode. The red-shifted/blue-shifted Raman frequency of the $A_1$ mode is attributed to the decreased/increased equilateral triangle projection of the chain on basal plane perpendicular to the *c*-axis, thus manipulating the basal plane vibration frequency and leading to Raman shift of $A_1$ mode.

**Zero strain-response along van der Waals *a*-axis.** To apply strain along the *a*-axis, the Te sample substrate was rotated by 90°, and the Raman spectroscopy was introduced along the *c*-axis on the same sample. The corresponding Raman spectra of *a*-axis strained Te are presented in Figure 2(d), where the $E^1$ mode has disappeared due to the optical anisotropic property reported previously that $E^1$ mode maximizes when Raman polarization is along *a*-axis, and vanishes when it is along *c*-axis[19]. Figure 2(e) and (f) illustrate the $A_1$ and $E^2$ peak positions as a function of tensile and compressive *a*-axis strains. Interestingly, both $A_1$ and $E^2$ have no response with various tensile and compressive strains, and the $A_1$ and $E^2$ Raman peak positions are independent of applied strains. The response along *a*-axis strain is totally different from that along *c*-axis, providing strong evidences of its anisotropic behavior. More importantly, the significant strain response along *c*-axis, together with no response along *a*-axis has conclusively verified the 1D van der Waals nature of Te thin film. The pronounced strain response when the deformation is applied along *c*-axis is directly associated with interatomic interactions between 1D covalent Te-Te bonds, while the absence of strain response in *a*-axis strains demonstrates no covalent bond, but pure weak van der Waals interaction in 2D plane of Te thin films.

**Magneto-transport property of Te in $B_z$ direction.** Several literatures have previously studied the anisotropic electrical behaviors of bulk Te, showing that resistivities of intrinsic Te at room temperature are $\rho_{||}$ = 0.26 Ω·cm and $\rho\perp$ = 0.51 Ω·cm, parallel and perpendicular to the *c*-axis, respectively[11]. However, the anisotropic magneto-transport properties of synthesized 1D van der Waals Te thin film have never been reported. The anisotropic magneto-transport properties with their temperature dependence are presented below. In the first experiment, the applied magnetic field is in *z*-direction and perpendicular to the *x-y* thin film plane. In the second experiment, the magnetic field is in *y*-direction along the *a*-axis of Te thin film. In the third experiment, the magnetic field is aligned to 1D Te atomic chains, *c*-axis, which is defined as the *x*-direction (Fig. 3b).

We first present the magneto-transport results of Te in $B_z$ direction, in which the magnetic field is applied perpendicularly to the 2D thin film plane. Magneto-resistance ($R_{xx}$) measurements of a typical Te thin film at different temperatures from 0.4 K to 5.0 K are presented in Figure 3(a). $R_{xx}$ exhibits a characteristic dip at small magnetic field regime, and this phenomenon is identified as WAL, indicating an existence of strong spin-orbit interaction in Te originated from its high atomic number and large electronic polarizability[23,24]. The WAL effect comes from a system with spin-orbit coupling in which the spin of a carrier is coupled to its orbit momentum. The spin of the carrier rotates as its goes around a self-intersecting path, and the direction of this rotation is opposite for the two branches of the loop. Because of this, the two paths along any loop interfere destructively, which leads to a lower net resistivity[26-30]. The importance of WAL is based on the consequence of spin-momentum locking and the full suppression of backscattering, resulting in a relative $\pi$ Berry phase acquired by carriers executing time-reversed paths. WAL is not an effect possessed only by Dirac materials, such as topological insulators and graphene, but a generic characteristic of the materials with strong spin-orbit coupling. WAL is temperature dependent as shown in Figure 3(a). We measured more than 10 devices with different film thickness, charge density and mobility. All show WAL characteristic. We use another sample with a similar thickness of 27.0 nm but more pronounced WAL features for the quantitatively analysis presented here (Figure 3d). As expected, the effect is strongly temperature-dependent and shows a deep peak at base temperature of 0.4 K. In addition, the effect diminishes when the temperature increases to 10 K, at which the localization is suppressed due to the decreased phase coherence length and spin-orbit scattering length at higher temperatures. We consider Hikami-Larkin-Nagaoka (HLN)[30] formula in the presence of spin-orbit coupling to discuss the WAL effect in $B_z$ field. We fit the low-field portion of the magneto-conductance curves using the HLN model,

$$\Delta\sigma(B) = \sigma(B) - \sigma(B=0) = n_s n_v (F(\frac{B}{B_\phi + B_{so}}) - \frac{1}{n_s}(F(\frac{B}{B_\phi}) - F(\frac{B}{B_\phi + 2B_{so}})))$$

$$F(Z) = \psi(\frac{1}{2} + \frac{1}{Z}) + ln(Z), B_\phi = \frac{\hbar}{4eL_\phi^2}, B_{so} = \frac{\hbar}{4eL_{so}^2}$$

where $\psi$ is the digamma function, $L_\phi$ is the phase coherence length, $L_{so}$ is the spin-orbit scattering length, $e$ is the electronic charge, $h$ is the Planck's constant, $B$ is the magnetic field, and $n_s$ and $n_v$ are spin degeneracy and valley degeneracy, respectively. Figure 3(e) shows the phase coherence length $L_\phi$ and the spin-orbit scattering length $L_{so}$ as a function of temperature. We should note that the maximum phase coherence length extracted from the small magnetic field portion is ~1.8 μm at 0.4 K at -30 V back gate bias. As we increase the temperature from 0.4 K to 10 K, the temperature dependence of the phase coherence length demonstrates a strong power-law behavior of $L_\phi \sim T^{-\gamma}$ with a power exponent $\gamma$ approximately 0.5. This value matches the observation from previous studies that electron-electron or hole-hole scattering would give $L_\phi$ proportional to $T^{-0.5}$ in a 2D system[31]. However, more refined experiments are needed to reveal its special transport property of 1D van der Waals structure in 2D thin film system when the magnetic field is perpendicular to the 2D plane. We find that $L_{so}$ is also temperature dependent, and the maximum spin-orbit scattering length at base temperature of 0.4 K is ~1.1 µm, and this number is promising for future Te-based spintronics applications. Temperature dependences of carrier concentration and mobility of Te are examined in this work as well, as depicted in Figure 3(f). The 2D carrier density $n_{2D}$ is determined from $n_{2D} = \frac{B}{e\rho_{xy}}$, with a magnitude of $1.8 \times 10^{13}$ cm$^{-2}$ in thin film Te is recorded at -30 V back gate bias and 0.4 K. Hall mobility, obtained from $\mu_H = \frac{L}{W}\frac{1}{R_{xx}n_{2D}e}$, where $L$ is the channel length, $W$ is the channel width, and $R_{xx}$ is the longitude resistance, has been measured with various temperatures. The Hall mobility remains constant at cryogenic temperatures with a maximum value of ~1,002 cm$^2$/Vs. We calculate the mean free path of Te thin films by applying equation,

$$l_m = \tau v_F$$

where $\tau$ is mean free time and $v_F$ is Fermi velocity. Related mobility and 2D carrier density are extracted from Figure 3(e) and 3(f). The estimated mean free path is ~ 700 nm at $\mu = 1,002$ cm$^2$/Vs and $n_{2D} = 1.8 \times 10^{13}$ cm$^{-2}$ which is comparable with our coherence length extracted from WAL effect. In our experiment, more than 10 samples have been systematically fabricated and measured, and a maximum mean free path of ~ 1 µm is obtained with a sample that has $\mu$ ~ 2000 cm$^2$/Vs and $n_{2D}$ ~ $8 \times 10^{12}$ cm$^{-2}$, which indicates a decent quality of our single crystal Te thin films.

**Magneto-transport property of Te in $B_y$ direction.** The $R_{xx}$ of the same sample with the magnetic field in $B_y$ direction is presented in Figure 4. The WAL effect is vanished with the applied $B_y$ magnetic field in the 2D film plane but perpendicular to the Te helical atomic chains. As shown in Figure 4, we do not observe any pronounced characteristics in magneto-resistance and Hall resistance depending on the applied external magnetic field in this configuration.

**Magneto-transport property of Te in $B_x$ direction.** The last and the most important experiment is the magneto-transport properties of Te thin film in $B_x$ direction, where the applied magnetic field is in line with 1D *c*-axis. Figure 5(a) shows the magneto-resistance of Te at different temperatures, in which the WAL effect is also observed within small magnetic field regime. In addition, UCF effect has been observed in our Te thin film at high fields. UCF is a quantum interference effect of diffusive charge carriers, observed commonly in the mesoscopic systems of semiconductors, metals, and 2D Dirac graphene. The mesoscopic nature or finite size of a weakly disordered sample results in the loss of the self-averaging of its physical properties[32,33]. Applying a magnetic field in semiconductors varies the phase of the wave function of a charge carrier, where the magnitude and interval of conductance fluctuation are closely related to the phase coherence length. The UCF effect, as presented in Figure 5(c), is observed in our Te thin film sample within the moderate magnetic field and low temperature regime, and this phenomenon is robust, and persisting to increased temperature up to 2 K. The periodic conductance fluctuation patterns can be observed repeatedly, and they present the similar features in the different magneto-conductance curves measured at different temperatures. To quantitatively compare the measured UCF magnitudes with theoretical predictions, we have generated root-mean-square (rms) magnitude of the magneto-conductivity fluctuations,

$$\Delta G_{rms} = \sqrt{\langle [\Delta G(B) - \langle \Delta G(B) \rangle]^2 \rangle}$$

where ⟨...⟩ denotes an average over the magnetic field which is an ensemble average over impurity configurations. Extracted $\Delta G_{rms}$ is demonstrated in Figure 5(d) with different temperatures. UCF theory predicts a fluctuation magnitude of 0.73 $e^2/h$ for a weakly disordered, 1D mesoscopic wire at $T=0$ K and at zero magnetic field[31]. At $T>0$ K, we limit our discussion to the 1D regime where the coherence length is much shorter than the sample size $L$ (6 µm), and the Te chain can be considered as a series of uncorrelated segments of length $L_\phi^{UCF}$. The relation between fluctuations rms and $L_\phi^{UCF}$ can be described in 1D system as[33,34,35],

$$\Delta G_{rms} = \sqrt{12} \frac{e^2}{h} \left( \frac{L_\phi^{UCF}}{L} \right)^{3/2}$$

The obtained values of $L_\phi^{UCF}$ extracted from observed UCF amplitudes can be used to verify the anisotropic property by comparing with $L_\phi$ from WAL effect along $B_z$. As shown in Figure 5(e), the behavior of $L_\phi^{UCF}$ is significantly different from what we saw in Figure 3(e), and we interpret it as the result of the anisotropic transport in Te where 1D chain transport differs from 2D thin film transport. The extracted phase coherence lengths from UCF follow a power law dependence of temperature $L_\phi \sim T^{-0.33}$, consistent with the dephasing mechanism being the carrier-carrier collisions with small energy transfers as observed commonly in 1D nanowire systems with an exponent value of 0.33[34,35]. Finally, we discuss the pronounced temperature dependent features of the conductance near zero magnetic field, presented in Figure 5(a) as huge peaks in magneto-resistance. Due to the localization effect within the small magnetic fields, it is hard to determine the real conductance at exact zero magnetic field. Instead of reading the numbers at magnetic field equals to zero, we use peak conductance near ±0.5 T as $\sigma(B=0)$ which we believe this value is approximately the intrinsic zero-field magneto-conductance with localization effect excluded. For the maximum gate bias of -80 V, conductance $\sigma$ in 1D system is given by[26],

$$\sigma(T) = \sigma_0 - \frac{2e^2}{h} \frac{L_\phi}{L}$$

since we have found that $L_\phi \sim T^{-0.33}$, we can rearrange our equation by replacing with phase coherence length with temperature. We fitted our experimental data from 0.4 K up to 2.0 K with a constant Drude conductance, and theoretical prediction is shown in red solid line as depicted in Figure 5(f). The functional form of the fit is consistent with the experimental data within low temperature, since our discussions are in quantum coherent regime. This result further confirms the 1D transport nature of Te thin films when the applied magnetic field is aligned with the helical atomic chains.

**Conclusion**

In summary, we have experimentally demonstrated the 1D van der Waals structure of Te thin film synthesized recently by our liquid phase method. Mechanical strain experiments have been conducted to verify its anisotropic crystal structure, in which the strong intra-chain covalent bond gives rise to the significant strain response along *c*-axis, and the zero strain-response along *a*-axis is due to the van der Waals interaction between helical chains. In addition, magneto-transport of Te with three different magnetic configurations proves its anisotropic property, resulting in dramatically distinct magneto-resistance behaviors. Specifically, phase coherence length extracted from WAL effect, $L_\phi \sim T^{-0.5}$ claims its 2D transport behavior when *B*-field is perpendicular to *x-y* plane. In contrast, $L_\phi \sim T^{-0.33}$ is obtained from UCF when the magnetic field is along *c*-axis of Te, confirming its nature of 1D transport mechanism.

**Methods**

**Sample synthesis.** The samples were grown through the reduction of sodium tellurite ($Na_2TeO_3$) by hydrazine hydrate ($N_2H_4$) in an alkaline solution at temperatures from 160-200 ºC, with the presence of crystal-face-blocking ligand polyvinylpyrrolidone (PVP). The amount of hydrazine is maintained at a high level to ensure the complete reduction of $TeO_3^{2-}$ into Te. A controlled, slow release of the Te source was achieved under the alkaline condition for maintaining the Te concentration, critical for the kinetic control of the 2D growth. The Te thin films were then transferred onto a $SiO_2$/Si substrate after a post-synthesis Langmuir-Blodgett assembly process, and been used for fabrication process.

**Raman measurements.** All Raman measurements were carried out on a HORIBA LabRAM HR800 Raman spectrometer. The system is equipped with a He-Ne excitation laser (wavelength 632.8 nm), an 1800 g/mm grating and a Nikon ×50 (NA = 0.45) long-working-distance objective lens. Subsequent Raman spectroscopy of strained Te studies was performed under an excitation laser power of 0.17 mW, sufficiently low to avoid excessive sample heating.

**Te thin film device fabrication.** Thin film Te with a similar thickness as compared to strain experiment was transferred from solution to a heavily doped silicon substrate with a $SiO_2$ capping layer. Hall bars were defined by the e-beam lithography, followed by Ni/Au contacts, deposited via e-beam evaporation. A magnitude of 100 nA input current was sent from the source to the drain, which is large enough to minimize measurement noise, but not too much to induce current heating at low temperature. Transport measurements were carried out in a $He^3$ cryostat with a superconducting magnet using a Stanford Research 830 lock-in amplifier.

FIGURES

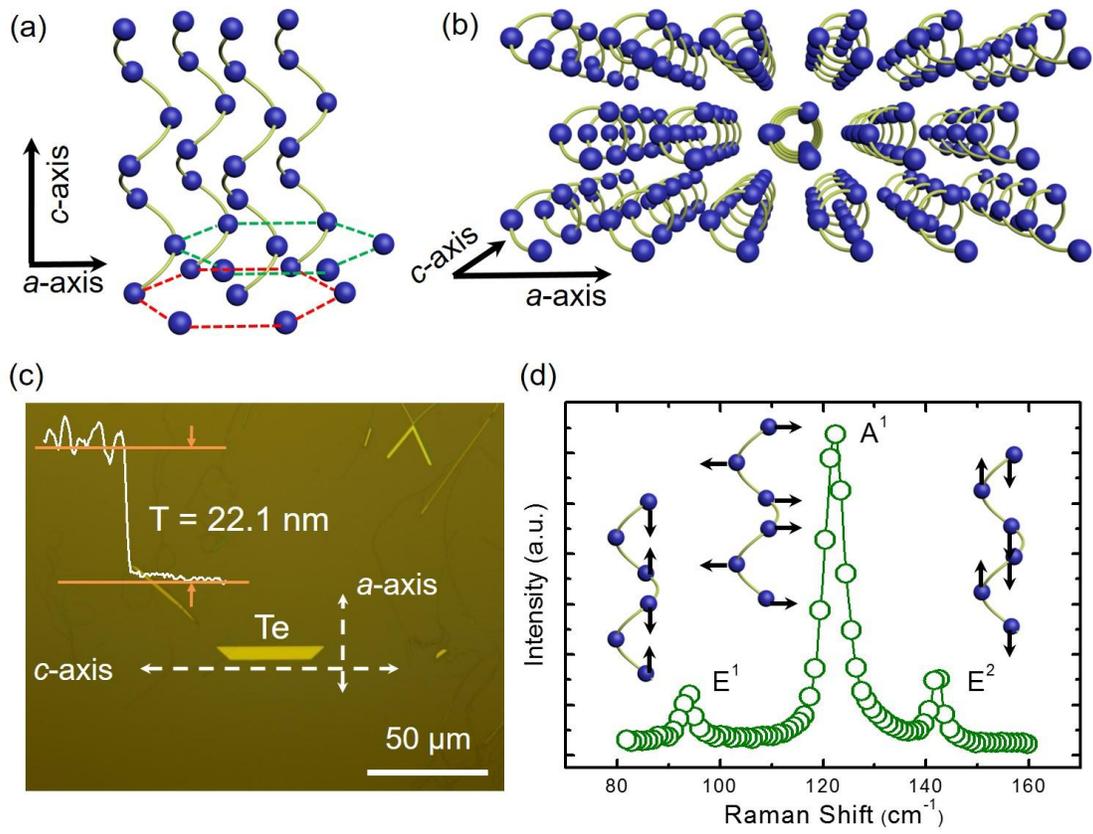

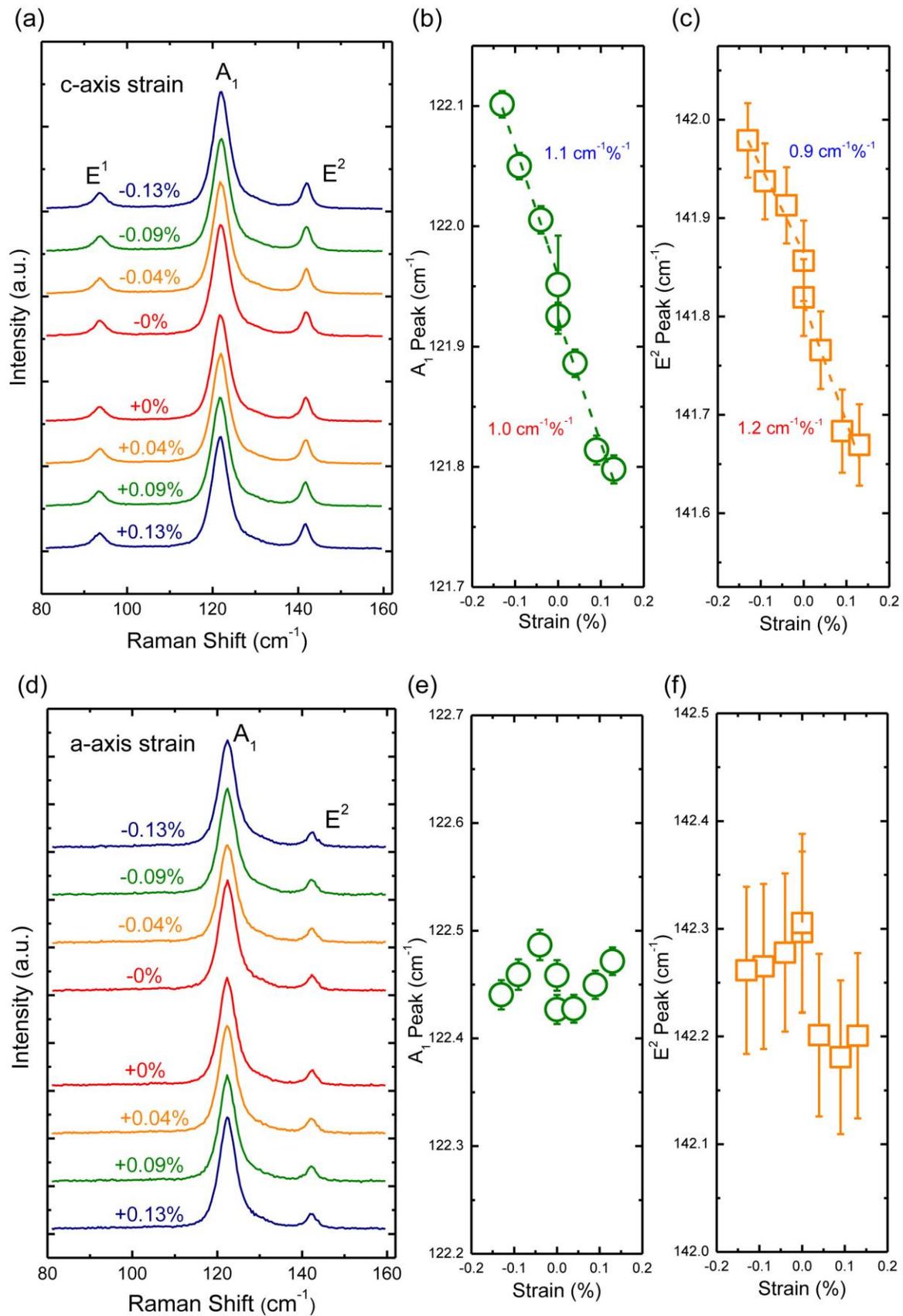

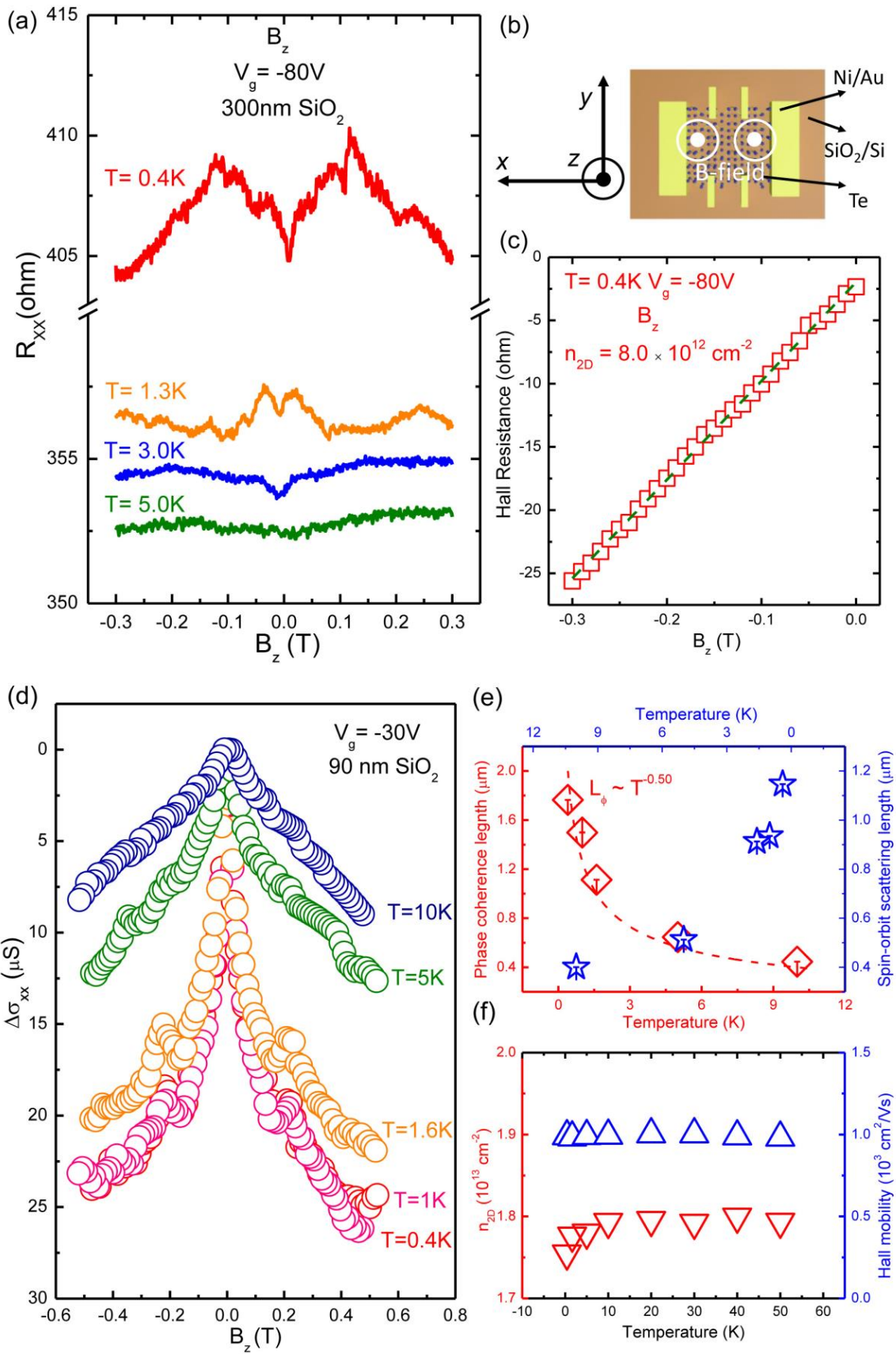

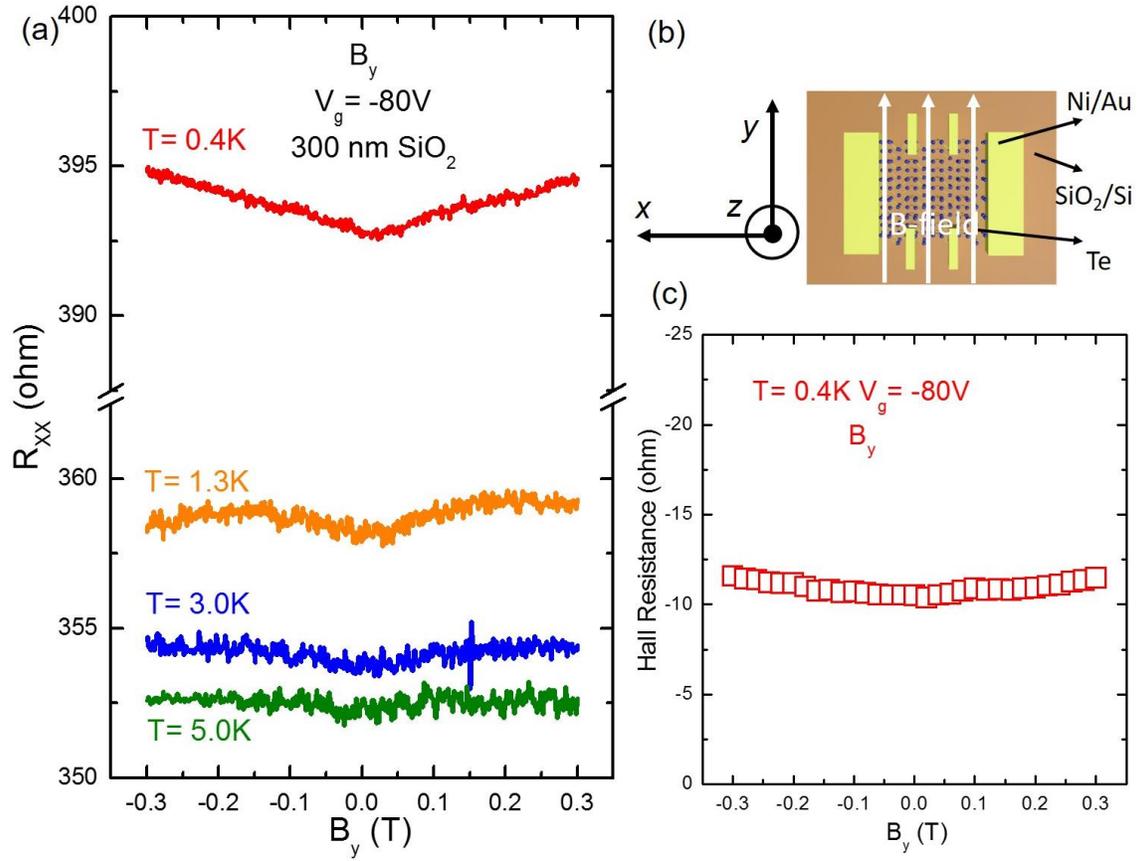

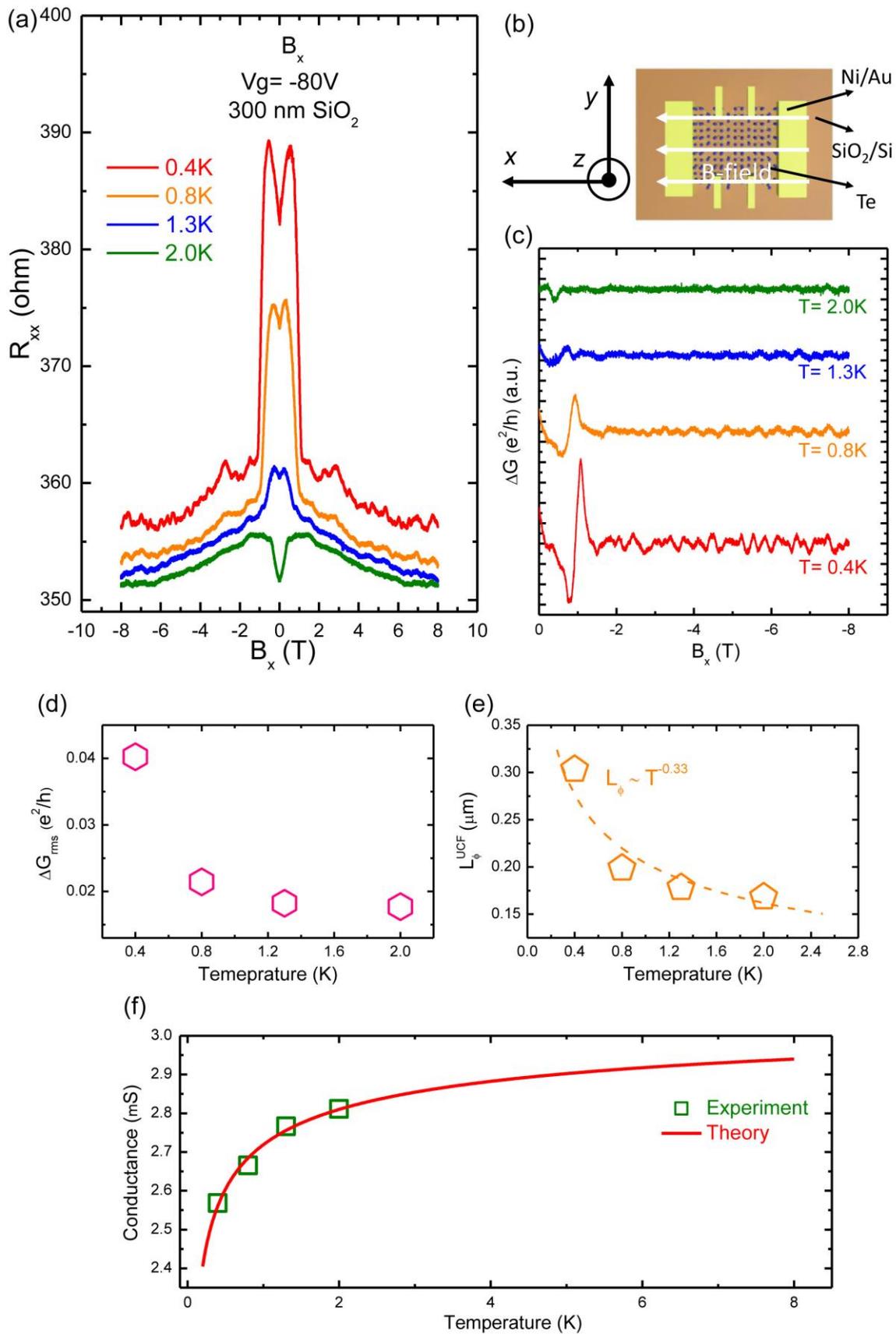

**Figure 1 | Te characterization.** (a) (b) Lattice structure of Te. The packing of the chains is determined by the green and red dashed lines, which represent pure van der Waals force in the basal plane. The solid yellow lines designate the primary covalent bonds. Each Te atom shares covalent bonds with its two nearest neighbors in the chain. Every fourth atom is directly located above another atom, that the projection of the chain on a plane perpendicular to the chain direction is an equilateral triangle. (c) Optical image of the 22.1 nm thick Te thin film with two principle lattice orientations. Scale bar in optical image is 50 µm. (d) Atomic vibrational patterns of $E^1$, $A_1$, and $E^2$ phonon modes in Te Raman spectra, and their corresponding atom vibration directions.

**Figure 2 | Raman evolution of uniaxial strained Te.** (a) Raman spectra of Te thin film for both tensile and compressive $c$-axis strains. Raman shift of (b) $A_1$ and (c) $E^2$ modes in $c$-axis strained Te. (d) Raman spectra of Te thin film for both tensile and compressive $a$-axis strains. Raman shift of (e) $A_1$ and (f) $E^2$ modes in $a$-axis strained Te. The dashed lines show linear fit results, and error bars are determined from Lorentzian peak fittings.

**Figure 3 | Magneto-transport of Te in $B_z$ direction.** (a) Magneto-resistance of Te thin film with varied $B_z$ field for different temperatures. The sample has a back-gate bias of -80 V, where the $SiO_2$ thickness is 300 nm. (b) Device scheme configuration and magnetic field indicator. (c) Hall effect of Te thin film with varied $B_z$ field at base temperature of 0.4 K, and -80 V. (d) A different sample that has a pronounced WAL effect. The WAL effect has been measured from 0.4 K, up to 10 K. The back-gate bias is -30 V, in which the back-gate dielectric is 90 nm instead. (d) Phase coherence length and spin-orbit scattering length extracted from WAL effect vary with measured temperatures. The phase coherence length has demonstrated a strong power-law behavior, and dashed line indicates theoretical simulation $L_\phi \sim T^{-\gamma}$ with a power exponent $\gamma =$

0.50. Our experimental results match the theoretical prediction that carrier scattering would give $L_\phi$ proportional to $T^{-0.5}$ in the 2D system. Error bars are determined from HNL fittings. (f) 2D carrier density and Hall mobility as a function of temperatures.

**Figure 4 | Magneto-transport of Te in $B_y$ direction.** (a) Magneto-resistance of Te thin film with varied $B_y$ field for different temperatures. The sample has a back-gate bias of -80 V, where the $SiO_2$ thickness is 300 nm. (b) Device scheme configuration and magnetic field indicator. (c) Hall effect of Te thin film with varied $B_y$ field at base temperature of 0.4 K, and -80 V.

**Figure 5 | Magneto-transport of Te in $B_x$ direction.** (a) Magneto-resistance of Te thin film with varied $B_x$ field for different temperatures. The sample has a back-gate bias of -80 V, where the $SiO_2$ thickness is 300 nm. (b) Device scheme configuration and magnetic field indicator. (c) $\Delta G(T)$ curves with varied $B_x$ field. The UCF effect is robust, and persisting to increased temperature up to 2 K. (d) $\Delta G_{rms}$ changes with different measured temperatures. (e) The phase coherence length has demonstrated a strong power-law behavior, and dashed line indicates theoretical simulation $L_\phi \sim T^{-\gamma}$ with a power exponent $\gamma = 0.33$. Our experimental results match the theoretical prediction that carrier scattering would give $L_\phi$ proportional to $T^{-0.33}$ in the 1D system. (f) Temperature dependence of the conductance at zero magnetic field. The red solid line is theoretical prediction, and green squares are experimental data.

ACKNOWLEDGEMENTS

The authors would like to thank Prof. Leonid Rokhinson and Zhong Wan for valuable discussions and experiment support. The authors would also like to thank Tim Murphy, Ju-Hyun Park, Glover Jones, and Hongwoo Baek at National High Magnetic Field Laboratory for technical assistances.

SUPPORTING INFORMATION

The supporting information is available free of charge on the ACS Publication website. Additional details for repeatibility of strain experiment, air-stability of Raman spectra, laser polarization configurations, magneto-resistance of $B_z$ direction in large B field, Hall effect measuremtns of $B_z$ direction, and gate dependent measurements of $B_x$ direction are in the suppproitng information.

AUTHOR CONTRIBUTIONS

P.D.Y. and W.W. conceived the idea, designed and supervised the experiments. Y.D. performed the strain experiments and analyzed the experimental data. Y.D., G.Q., and M.S. performed magneto-transport experiments and analyzed the experimental data. Y.W. synthesized Te thin films. X.X. supported Raman spectra measurements. Y.D., G.Q., Y.W., M.S., X.X., W.W., and P.D.Y. co-wrote the manuscript.

COMPETING FINANCIAL INTERESTS STATEMENT

The authors declare no competing financial interests.

Supplementary Information for:

# 1D van der Waals Material Tellurium: Raman Spectroscopy under Strain and Magneto-transport

Yuchen Du[1,4], Gang Qiu[1,4], Yixiu Wang[2], Mengwei Si[1,4], Xianfan Xu[3,4], Wenzhuo Wu[2,*], and Peide D. Ye[1,4,*]

[1] School of Electrical and Computer Engineering, Purdue University, West Lafayette, Indiana 47907, United States

[2] School of Industrial Engineering, Purdue University, West Lafayette, Indiana 47907, United States

[2] School of Mechanical Engineering, Purdue University, West Lafayette, Indiana 47907, United States

[3] Birck Nanotechnology Center, Purdue University, West Lafayette, Indiana 47907, United States

* Address correspondence to: yep@purdue.edu or wenzhuowu@purdue.edu

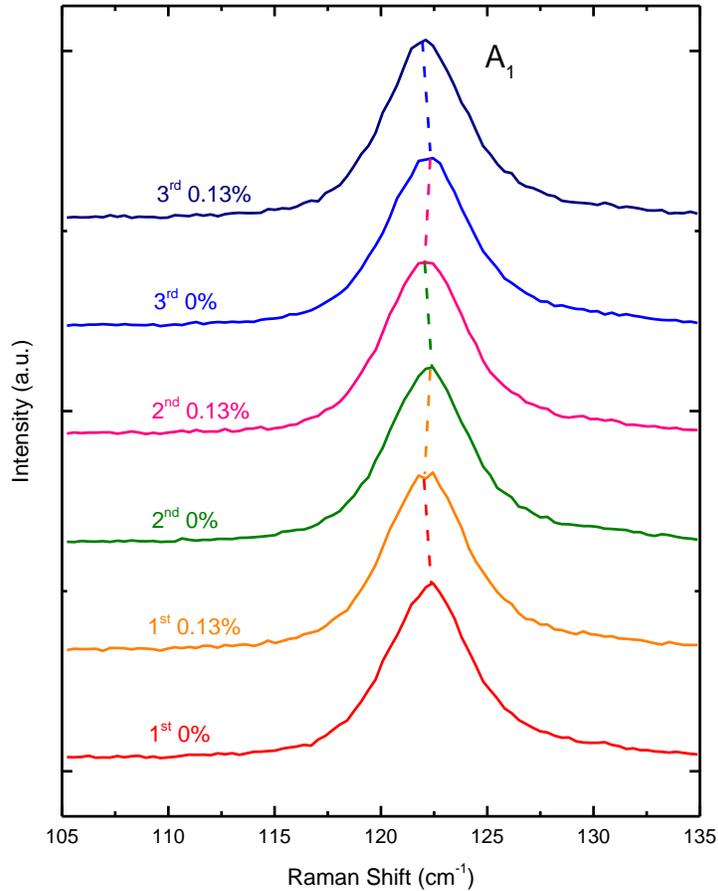

**Supplementary Figure 1 | Multiple loading and unloading processes from Raman spectroscopy measurement.** Raman spectra of uniaxial tensile strained Te for multiple loading and unloading processes.

**Supplementary Note 1 | Multiple loading/unloading processes**

Multiply loading and unloading processes have been conducted to verify the reversibility of Te with respect to strain measurements. The $A_1$ mode of Te has been chosen to monitor the peak position variation with different strains. $A_1$ peak at ~122 cm$^{-1}$ experiences a red-shift once the 0.13% tensile strain is applied, and restores back to original position when strain is released. The red-shift of the active mode under strain as well as the blue-shift back to original position due to strain relaxation are clearly observed in our experiment, indicating a good strain reversibility of Te.

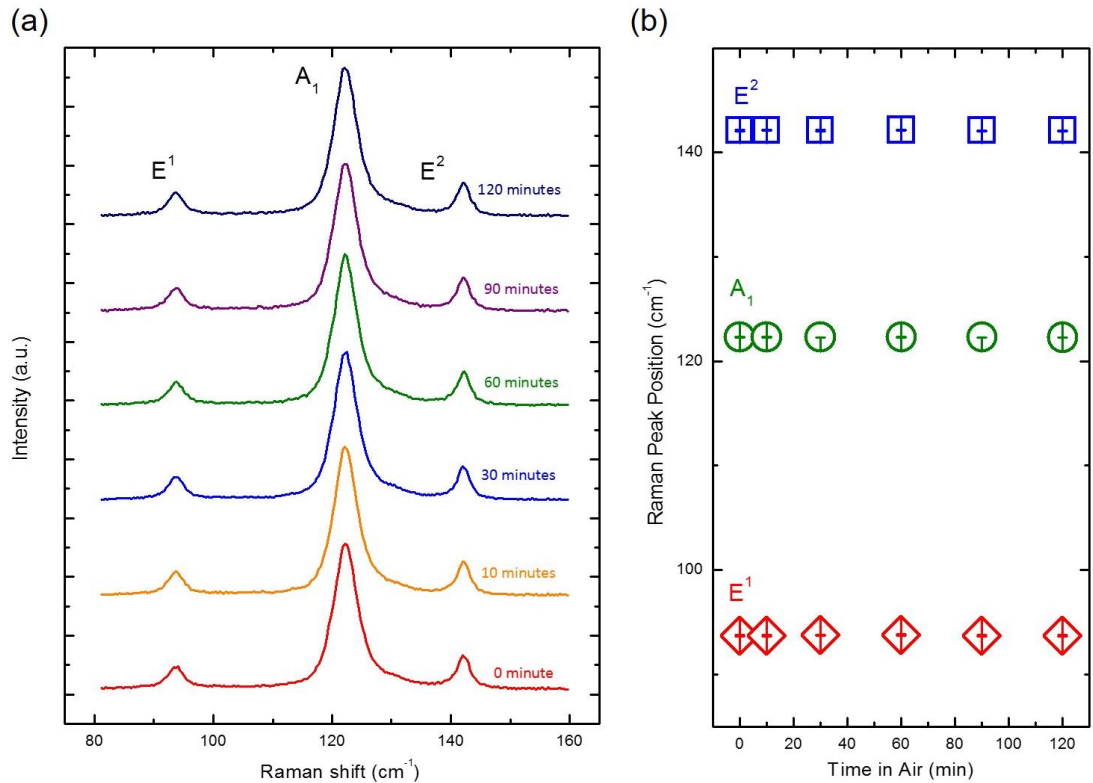

**Supplementary Figure 2 | Air-stability studies of Te Raman spectroscopy measurement.** (a) The evolution of Raman intensity as a function of time up to 120 minutes. The peak positions of (b) $E^1$, $A_1$, and $E^2$ modes for various time.

**Supplementary Note 2 | Air-stability of Raman spectra**

Air-stability of Te has been monitored through measuring the Raman spectra as a function of time up to 120 minutes. As shown in Supplementary Figure S2, the Raman spectra of Te thin film demonstrate a very stable characteristic within a measured time frame. For each vibration mode, the peak position does not show a systematical shift, which indicates Raman shift of uniaxially strained Te is directly associated with the strain induced interatomic vibration.

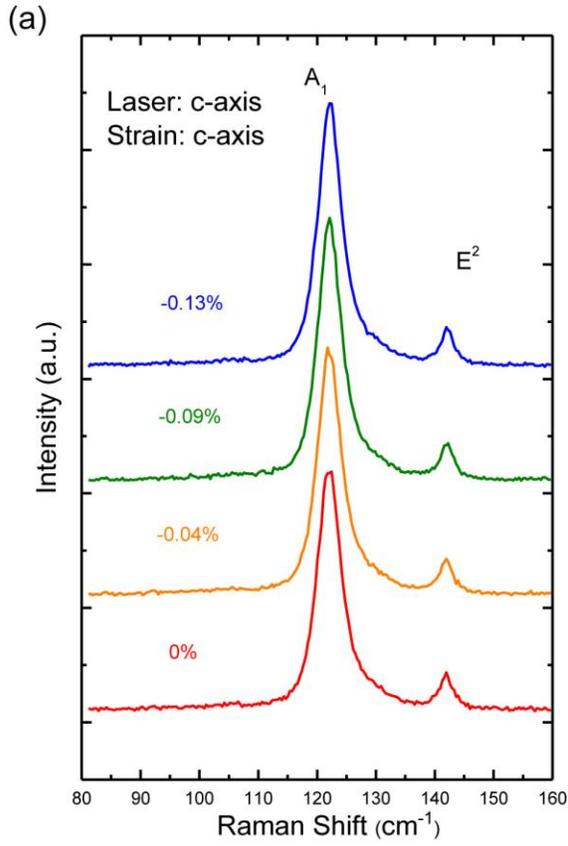
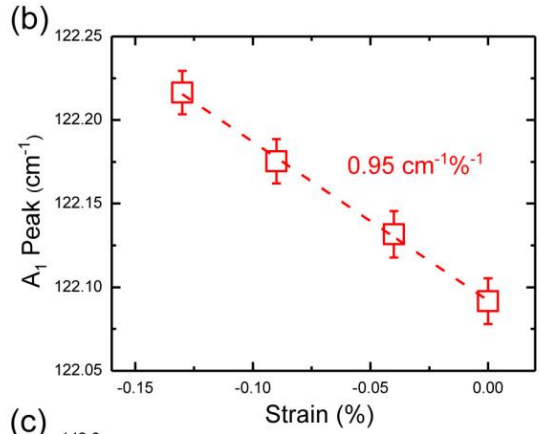
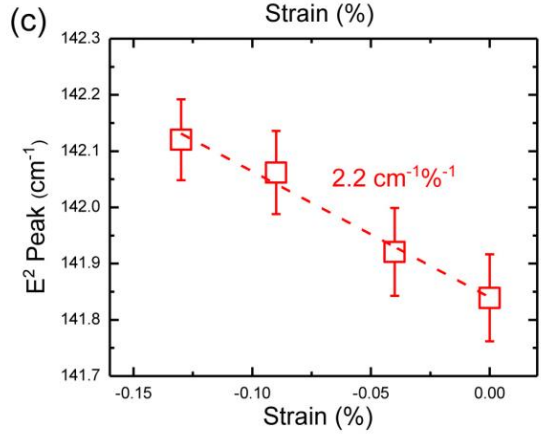
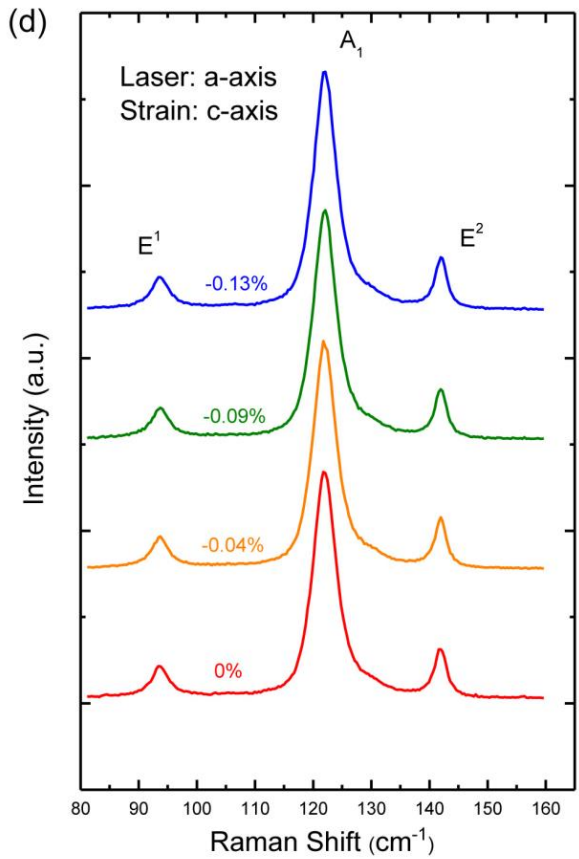
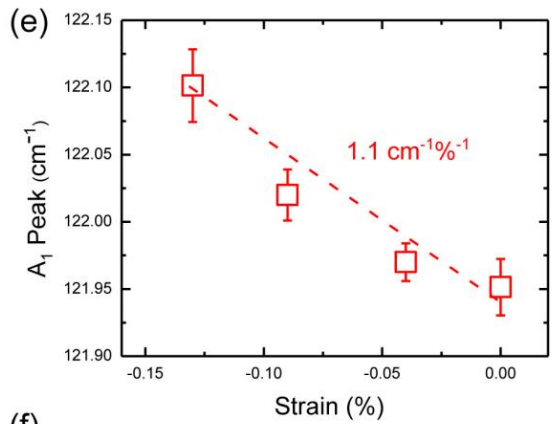
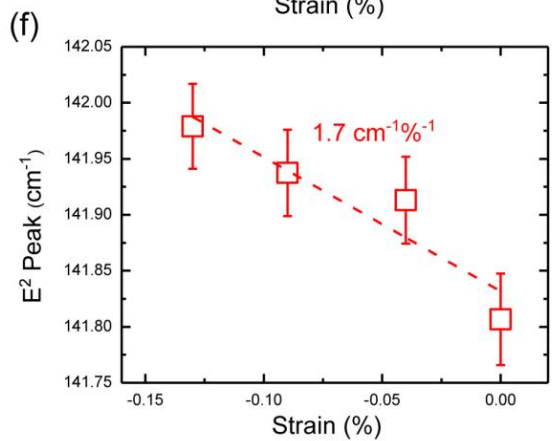

**Supplementary Figure 3 | Laser polarization configurations.** (a) The evolution of Raman intensity for *c*-axis laser and *c*-axis strain. The peak positions of (b) $A_1$ and (c) $E^2$ modes. (d) The evolution of Raman intensity for *a*-axis laser and *c*-axis strain. The peak positions of (e) $A_1$ and (f) $E^2$ modes.

**Supplementary Note 3 | Laser polarization configurations.**

For the results shown in main manuscript, the strain direction is perpendicular to the laser polarization direction based on our apparatus set-up, *i.e.* *c*-axis strain is recorded by performing *a*-axis Raman spectra, and *a*-axis strain is recorded by performing *c*-axis Raman spectra. To further confirm the 1D van der Waals nature of Te, we have conducted another experiment to exclude any deviation that may be introduced by laser polarization configuration. For two strain experiments with the same strain direction (*c*-axis), we separately measured the Raman spectra with respect to different laser polarizations. We found out that the direction and magnitude of Raman shifts are very similar, and we can confirm that there exists no laser polarization induced deviation in our experiment.

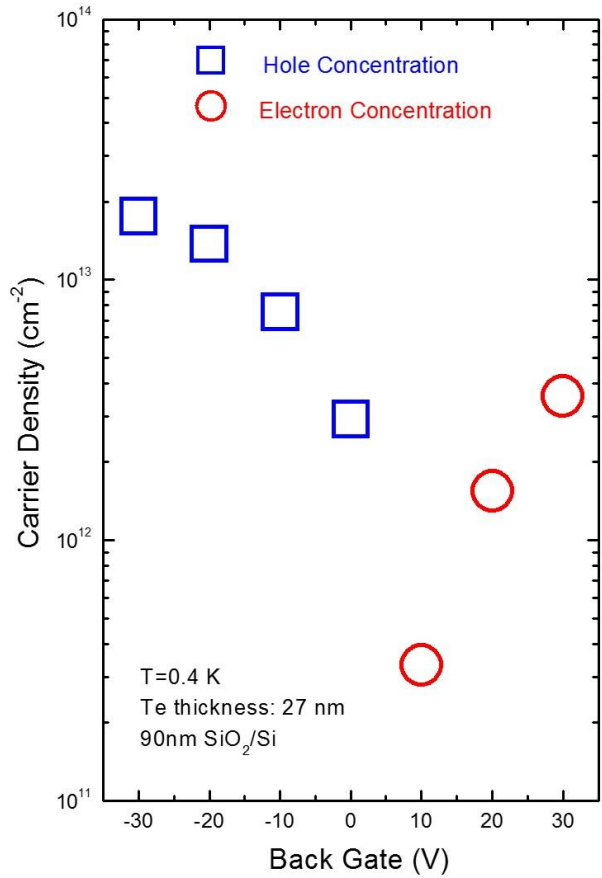

**Supplementary Figure 4 | Carrier density of $B_z$ direction with different back-gate bias.** The carrier density can be tuned from hole to electron by applying different back-gate bias. 2D carrier density with different back-gate bias at base temperature of 0.4 K.

**Supplementary Note 4 | Carrier density of $B_z$ direction with different back-gate bias.**

The 2D carrier density can be tuned from hole to electron by applying different back-gate bias. The intrinsic carrier concentration at zero gate bias in this 27.0 nm Te sample is $2.9 \times 10^{12}$ cm$^{-2}$.